# Dielectric-barrier discharges in two-dimensional lattice potentials

J. Sinclair and M. Walhout
*Department of Physics & Astronomy, Calvin College, Grand Rapids, MI 49546, USA*
(Submitted to and received by Physical Review Letters, 30 August 2011)

We use a pin-grid electrode to introduce a corrugated electrical potential into a planar dielectric-barrier discharge (DBD) system, so that the amplitude of the applied electric field has the profile of a two-dimensional square lattice. The lattice potential provides a template for the spatial distribution of plasma filaments in the system and has pronounced effects on the patterns that can form. The positions at which filaments become localized within the lattice unit cell vary with the width of the discharge gap. The patterns that appear when filaments either overfill or under-fill the lattice are reminiscent of those observed in other physical systems involving 2d lattices. We suggest that the connection between lattice-driven DBDs and other areas of physics may benefit from the further development of models that treat plasma filaments as interacting particles.

PACS number(s): 47.54.-r, 52.80.Tn, 63.90.+t, 64.60.My, 89.75.Kd

This Letter connects the study of pattern formation in dielectric-barrier discharge (DBD) systems with areas of physics involving particles and interactions in two-dimensional (2d) lattice potentials. To date there have been many investigations of how the plasma filaments in a DBD can arrange themselves in various 1d and 2d patterns in a lateral plane perpendicular to the applied, oscillating electric field. In all of that work, the applied field has been uniform in the lateral plane, and "spontaneous" pattern formation has been governed almost entirely by the mutual interactions between filaments. In the experiments to be presented here, we apply a non-uniform field with an amplitude profile having the form of a 2d square lattice.

The lattice potential provides a template for DBD filament patterns, just as a crystalline substrate can do for the distribution of adatoms on a solid surface. Additionally, like optical lattices used to hold arrays of ultracold atoms, our DBD lattice has an adjustable binding strength and allows for the lattice period to be changed easily. In fact, our system of plasma filaments can be understood as an array of electric dipoles confined to a plane, repelling each other and possibly hopping between lattice sites. From this perspective it is similar to a system of fermionic dipolar molecules in a square optical lattice, for which patterned phases were recently identified [1].

We preface the discussion of our experiments with a few comments about conventional DBD systems. The ac plasma generated by a high-pressure DBD is filamentary; each half-cycle of the driving voltage generates an array of short-lived plasma filaments distributed in the plane perpendicular to the applied electric field. The process leaves surface charges, or "footprints," on the dielectric surfaces that cover the electrodes, and the deposited charges produce a memory effect in the system, so that each filament array tends to coincide spatially with that of the previous half-cycle. The result is a quasi-stable filament array, firing thousands of times per second, in which visible patterns may form spontaneously [2]. Spatially and temporally resolved measurements have revealed complex structures underlying what is seen with the naked eye. It is now known, for instance, that filament arrays can ignite in a sequence of distinct stages during a single half-cycle, with each stage producing a sub-component of the integrated visible pattern [3, 4]. Progress has also been made in measuring and modeling charge-footprint distributions, and promising steps have been taken toward understanding the interactions between filaments [5-9]. All of this work has contributed significantly to a rich conceptual framework that we now bring to our present investigation of DBD systems driven by lattice potentials.

Our DBD chamber is composed of three parts: a lower cartridge filled with a water electrode, an upper cartridge containing a "pin-grid" electrode, and a skirt structure (Fig. 1). The upper and lower cartridges slip into the skirt with an average clearance of less than 0.1 mm around the edges. The discharge gap is the vertical space between the two the opposing cartridge faces, which are made from square glass plates with side length 16 cm and thickness $1.50 \pm 0.02$ mm. An array of inlet holes in the skirt delivers a mixture of argon and helium into the gap. The gas flows through the system at a slight overpressure and escapes through the leaks at the cartridge edges.

Except for the two glass plates, which serve as our dielectric barriers, all the walls of the cartridges and the skirt are made of transparent polycarbonate plastic. This construction allows us to look into the gap from any position along its perimeter. In addition, the lower cartridge is placed on a transparent plastic shelf, so that we can look through the water electrode from below and observe the full lateral extent of the discharge region. We use a high-definition video camera to capture images from this vantage point.

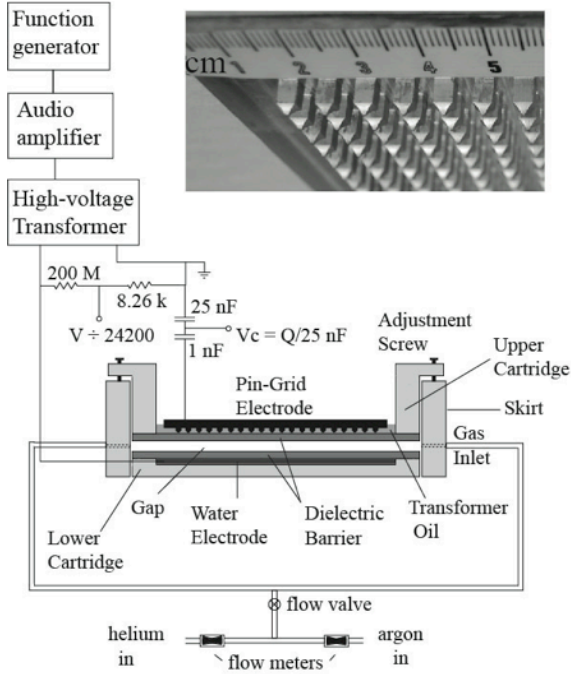

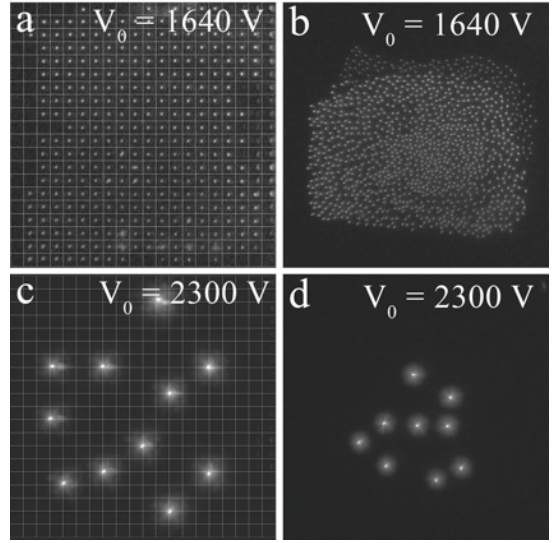

FIG. 1. Diagram of the experimental setup and photograph of the pin-grid electrode plate. The plate and all 400 pins are fabricated from a single piece of aluminum. Along each of the main axes, the spacing between adjacent pins is 6.35 mm.

FIG. 2. Filament distributions in pure argon just after breakdown, with and without the lattice potential. For the lattice-driven cases (a, c) gridlines are drawn to produce square unit cells centered on the electrode pins. (a) With $w = 1$ mm, filaments in the lattice potential are strongly localized at the pin positions; (b) when a planar electrode is used, filaments crowd more closely together and exhibit much less spatial order. (c) For $w = 3$ mm, filaments in the lattice become localized at the corners of the unit cells, rather than at the centers, and their footprints develop multiple streamer extensions that reach toward nearby electrode pins. (d) By contrast, the footprints in the non-corrugated DBD exhibit no such azimuthal variation for $w = 3$ mm. The driving frequency is 13.5 kHz for each of the images.

Various electrode plates can be used in the upper cartridge. For this initial study, we select a pin-grid electrode that creates a 20×20 2d square lattice potential with a nearest-neighbor site spacing of 6.35 mm. The tips of the electrode pins rest directly on the top side of the upper glass plate and are immersed in a pool of transformer oil that ensures against unwanted pin-to-glass discharges. The upper cartridge sits on four screws that we can turn to adjust the gap width. To measure the gap width ($w$) we view the edges of the dielectric barriers through the skirt and visually align them with a handheld vernier caliper.

A sinusoidal high voltage ($V = V_0 \sin 2\pi\nu t$) is applied to the water electrode. Its frequency is set within the 10-20 kHz range and its amplitude is varied between 0 and 5 kV. We measure V by means of a resistive voltage divider circuit. The grid electrode is separated from the electrical ground by a series combination of two capacitors (1 nF and 25 nF). The system's electrode-glass-gap sandwich has a capacitance of less than 0.1 nF and is the element that limits the amount of charge ($Q$) stored on each capacitor. We monitor Q by measuring the voltage ($V_C$) across the 25-nF capacitor, which exhibits an amplitude of a few volts under normal operating conditions. We use a digital oscilloscope to record V and $V_C$ as functions of time.

Fig. 2 shows examples of filament distributions produced with (a, c) and without (b, d) the lattice potential. Each image is obtained just after breakdown in pure argon. For small gap widths ($w < 2.5$ mm), a tight-binding effect is observed at the site of each grid pin. Localization also happens for widths of 3-4 mm, but the stable positions shift from the pins to the corners of the lattice unit cell (Fig. 2c). Localization effects diminish steadily as the width is increased beyond 4 mm.

Whereas footprints in the non-corrugated potential remain cylindrically symmetric for large gap widths (> 3 mm), the ones in the lattice-driven DBD develop multiple streamer extensions that reach toward nearby pins (Fig. 2d). This effect is becomes more pronounced as the gap is widened. It is reasonable to assume that the charge patches left by each filament have comparable azimuthal structure and, therefore, that the interactions between filaments must be anisotropic. We will not study these interactions in detail here, but we suggest that such an effort could lead to fruitful connections with other areas of physics in which anisotropic interactions in 2d-lattices are a topic of active research [10].

After initial breakdown in the gas, the number of filaments increases as $V_0$ is raised. Eventually the lattice becomes overfilled, and filaments can form patterns with spatial frequencies higher than that of the underlying grid.

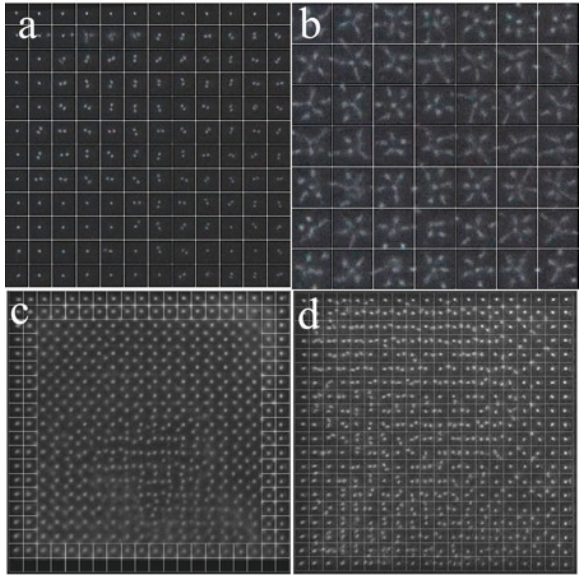

FIG. 3. Patterns on an overfilled lattice. (a) Argon discharge, $w = 1.4$ mm, $\nu = 13.5$ kHz, $V_0 = 2100$ V; (b) Argon, $w = 1.7$ mm, $\nu = 13.5$ kHz, $V_0 = 2570$ V; (c) Ar:He = 2:1, $w = 2$ mm, $\nu = 17.0$ kHz, $V_0 = 2400$ V; (d) Ar:He = 4:3, $w = 2$ mm, $\nu = 13.5$ kHz, $V_0 = 1890$ V. Multistage discharges occur in all four cases.

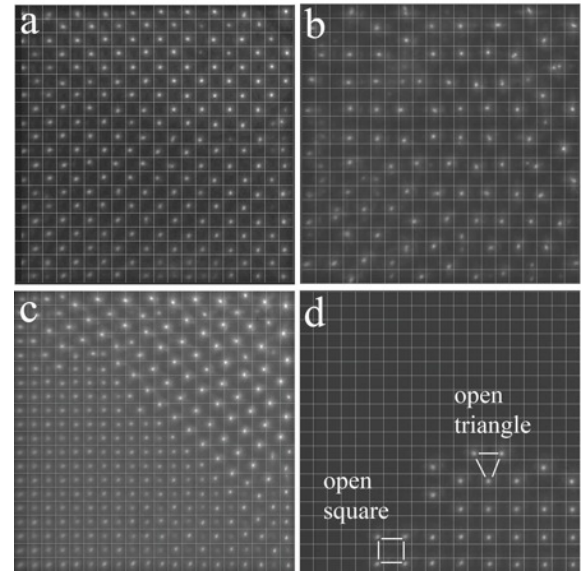

FIG. 4. Patterns on an under-filled lattice ($\nu = 13.5$ kHz). (a) A checkerboard pattern appears when Ar:He = 4:3, $w = 3.5$ mm, and $V_0 = 2530$ V; (b) A mixture of open squares and open triangles is seen in argon for $w = 3.0$ mm and $V_0 = 2680$ V; (c) When $V_0$ is raised to 6000 V in argon ($w = 3.5$ mm), new filaments form to create a filled domain over a large region of the lattice, while the remaining under-filled domain exhibits an altered spatial order. (d) When $V_0$ is decreased toward the discharge-sustaining level, many filaments turn off, leaving patches of under-filled patterns. In this instance, Ar:He = 4:3, $w = 3.5$ mm, and $V_0 = 1430$ V.

In pure argon discharges with small gap widths (Figs. 3a and 3b), individual filaments bifurcate into localized pairs, triplets and various star-like structures. While these "micro-patterns" are centered on individual lattice sites, there can be longer-range correlations between their individual azimuthal orientations.

In discharges like those shown in Figs. 3c and 3d for He-Ar mixtures, new filaments are spawned not through bifurcations at individual sites, but by new breakdown events in the spaces between the established, pin-centered filaments. The onset of interstitial filaments is accompanied by the appearance of an additional, pronounced step in the $V_C$ (capacitor-charging) signal. This suggests that the interstitials ignite during a distinct discharge stage that occurs at an otherwise quiescent phase of the driving oscillation. The time-averaged pattern is thus a superposition of multiple sub-patterns produced by a sequence of temporally separated stages. As mentioned previously, this kind of spatiotemporal pattern-formation scenario has been observed and partially explained in the context of conventional DBDs. However, our new setup imposes a spatial template causing the system to favor patterns commensurate with the underlying lattice.

The interstitial filaments in overfilled lattices can stabilize at characteristic positions within the unit cell, so that either of two distinct pattern types may appear. As seen in the top half of Fig. 3c, the pattern of second-stage filaments may be isomorphic with the lattice but shifted on the unit grid by a displacement vector of (½, ½). Each interstitial filament in this pattern is located on a corner of a unit cell of the lattice. In the other common pattern, which appears in small regions of Fig. 3c and in much of 3d, the interstitials line up along the grid axes connecting nearest-neighbor pins. On long time scales (~1 s) they can move along the axes, hop between axes, and push on each other. The motion of interstitial filaments in this pattern resembles that of molecules in smectic phases of liquid crystals [11] and merits further investigation.

Fig. 4 shows patterns in an under-filled lattice. These patterns tend to appear when the gap width is relatively large (~ 3mm), in which case the size of filament footprints is comparable to the lattice spacing. Figs. 4a and 4b show multi-stage discharge patterns that spread across the plane but do not fill each lattice site, despite the fact that $V_0$ is well above the usual breakdown voltage. The cause of this behavior lies in each filament's ability to inhibit breakdown events over an area extending beyond the nearest-neighbor sites. Note, however, that at very high values of $V_0$, the applied field can overcome the inhibition, so that new filaments can ignite in the empty sites. As seen in Fig. 4c, the high field can produce this lattice-filling effect over much of the discharge plane while also altering the spatial order in the remaining under-filled domain. We note that our setup does not allow us to tell whether individual filament images in Figs. 4a-c are produced by multiple (rather than single) discharge events occurring within each voltage half-cycle.

As $V_0$ is lowered toward the discharge-sustaining voltage, filaments begin to turn off, and the system is eventually left with under-filled patterned domains like the one shown in Fig. 4d. Our $V_C$ signal reveals only one discharge stage per voltage half-cycle in such situations, so we conclude that a single filament array (no superposition) is responsible for any patterns that appear. These patterns contain information about the interactions between filaments; this point can be understood as follows. Every filament must lie outside the breakdown-inhibition zone of every other filament. However, each "hole" or empty site should lie within an inhibition zone. Thus, the separation between nearest-neighbor filaments in any pattern gives an upper limit for the range of the breakdown-inhibition effect.

By repeatedly varying $V_0$ in order to fill and then deplete the lattice, we are able to observe stable single-stage patterns characterized by checkerboard, open-triangle, and open-square geometries. These patterns are well known in other square-lattice systems. They correspond, for instance, to phases that were recently predicted for ultracold dipolar molecules in under-filled optical lattices [1, 12]. We also observe other patterns that are only metastable because filaments within the array can hop from site to site, with a characteristic hopping time of about 1 s. One of these is the "knight-move" pattern, which was discussed recently in the context of surface phase transitions [13] and also appears in [1] with the label "5C."

In most areas of physics, lattice-based patterns are analyzed in terms of the motions and interactions of particles in a corrugated potential. It seems likely that the analysis of lattice-driven DBD dynamics, too, could benefit from a fully developed particle model of discharge filaments—that is, a model in which each filament is treated as a single particle. Studies of conventional DBDs have already begun to delineate the particle-like properties of filaments and their interactions [14], and now lattice-driven systems may serve as a new testing ground for further theoretical understanding. While filaments are not generally conserved and do not have well-defined masses, they do have mobility and can appear to attract or repel each other in particle-like ways. Moreover, it is worth considering that filament behavior on a lattice could be modeled using various lattice-related mathematical tools familiar in condensed-matter physics (e.g., creation and annihilation operators, hopping amplitudes, nearest-neighbor interactions, on-site repulsion potentials, as well as concepts related to cellular automata and percolation theory).

Our present study raises questions and possibilities that are new to the field of DBD studies but touch on longstanding knowledge in other areas. On the theoretical front, investigations of lattice-driven DBDs could both borrow from and contribute to the understanding of other pattern-forming, lattice-based systems. Filament patterns like those we observe are, in effect, "gaseous crystals," and it seems legitimate to ask whether they have properties analogous to conductivity, compressibility, and photonic band gaps [15]. Of course, there is also much more to learn about the kinds of filament patterns that can be produced in corrugated potentials. In the preliminary work presented here we have focused on a very limited region of parameter space, leaving the effects of different lattice symmetries, gas mixtures and ranges of driving frequency to be explored in future research.

JS thanks the Calvin College Physics and Astronomy Department for a John Van Zytveld Summer Research Fellowship. This work was also supported by NSF grant PHY-1068078.

[1] K. Mikelsons and J.K. Freericks, Phys. Rev. A **83**, 043609 (2011).
[2] U. Kogelschatz, IEEE Transactions on Plasma Science 30, 1400 (2002).
[3] M. Klein, N. Miller, and M. Walhout, Phys. Rev. E **64**, 026402 (2001).
[4] Lifang Dong, Weili Fan, Yafeng He, Fucheng Liu, Shufeng Li, Ruiling Gao, and Long Wang, Phys. Rev. E 73, 066206 (2006).
[5] Jiting Ouyang, Feng He, Shuo Feng, Zhinong Yu, Zhihu Liang, and Jianqi Wang, Appl. Phys. Lett. **89**, 031504 (2006).
[6] L. Stollenwerk, Sh. Amiranashvili, J.-P. Boeuf, and H.-G. Purwins, Eur. Phys. J. D **44**, 133 (2007).
[7] L. Stollenwerk, J.G. Laven, and H.-G. Purwins, Phys. Rev. Lett. **98**, 255001 (2007).
[8] L. Stollenwerk, New Journal of Physics **11**, 103034 (2009).
[9] L. Stollenwerk, Plasma Phys. Control. Fusion **52**, 124017 (2010).
[10] Y.-H. Chan, Y.-J. Han, and L.-M. Duan, Phys. Rev. A **82**, 053607 (2010).
[11] P.J. Collings and M. Hird, *Introduction to Liquid Crystals* (Taylor and Francis, Philadelphia, 1997).
[12] B. Capogrosso-Sansone, C. Tefzger, M. Lewenstein, P. Zoller, and G. Pupillo, Phys. Rev. Lett. **104**, 125301 (2010).
[13] A. Andrews, M. Novenstern, and L.D. Roelofs, ChemPhysChem **11**, 1476-1481 (2010).
[14] H.-G. Purwins, American Institute of Physics Conference Proceedings 993, PLASMA 2007, H.-J. Hartfuss, M. Dudeck, J. Musielok, and M.J. Sadowski, eds., pp. 67-74. (2008).
[15] Weili Fan, Xinchun Zhang, and Lifang Dong, Phys. Plasmas **17**, 113501 (2010). This paper examines photonic band-gap effects in conventional DBDs.